# Measurements of microjoule-level, few-femtosecond ultraviolet dispersive-wave pulses generated in gas-filled hollow capillary fibers

CHENG ZHANG,[1,†] TIANDAO CHEN,[2,3,†] JINYU PAN,[1,2,3] ZHIYUAN HUANG,[2,*] DONGHAN LIU,[2,3] DING WANG,[2] FEI YU,[1,4] DAKUN WU,[1] YU ZHENG,[7] RUOCHEN YIN,[7] XIN JIANG,[5,6] MENG PANG,[1,2,8] YUXIN LENG,[1,2,9] AND RUXIN LI[2]

[1]*Hangzhou Institute for Advanced Study, Chinese Academy of Sciences, Hangzhou 310024, China*
[2]*State Key Laboratory of High Field Laser Physics and CAS Center for Excellence in Ultra-intense Laser Science, Shanghai Institute of Optics and Fine Mechanics (SIOM), Chinese Academy of Sciences (CAS), Shanghai 201800, China*
[3]*Center of Materials Science and Optoelectronics Engineering, University of Chinese Academy of Sciences, Beijing 100049, China*
[4]*Key Laboratory of Materials for High Power Laser, Shanghai Institute of Optics and Fine Mechanics, Chinese Academy of Sciences, Shanghai 201800, China*
[5]*National Engineering Laboratory for Fiber Optic Sensing Technology, Wuhan University of Technology, Wuhan 430070, China*
[6]*Russell Division, Max Planck Institute for the Science of Light, Erlangen 91058, Germany*
[7]*IFiber (Ningbo) Optoelectronics Technology Co., LTD., Ningbo 315000, China*
[8]*e-mail: pangmeng@siom.ac.cn*
[9]*e-mail: lengyuxin@mail.siom.ac.cn*
*Corresponding author: huangzhiyuan@siom.ac.cn

**High-energy ultraviolet pulse generation in gas-filled hollow capillary fibers (HCFs) through dispersive-wave-emission process, has attracted considerable attentions in recent several years due to its great application potentials in ultraviolet spectroscopy and photochemistry. While the ability of this technique to deliver high-energy, ultrafast ultraviolet pulses is widely recognized, few-fs duration of μJ-level dispersive-wave (DW) pulses has, however, escaped direct measurements. In this letter, we demonstrate that using several chirped mirrors, few-fs ultraviolet DW pulses can be obtained in atmosphere environment without the use of vacuum chambers. The pulse temporal profiles, measured using a self-diffraction frequency-resolved optical gating set-up, exhibit full-width-half-maximum pulse widths of 9.6 fs at 384 nm and 9.4 fs at 430 nm, close to the Fourier-transform limits. Moreover, theoretical and experimental studies reveal the strong influences of driving pulse energy and HCF length on temporal width and shape of the measured DW pulses. The ultraviolet pulses obtained in atmosphere environment with μJ-level pulse energy, few-fs pulse width and broadband wavelength tunability are ready to be used in many applications.**

Dispersive wave (DW) emission in gas-filled hollow capillary fibers (HCFs) provides an elegant means of generating high-energy, wavelength-tunable light sources at short wavelengths [1-3]. When ultrafast driving pulses at near-infrared were launched in a section of gas-filled HCF, high-order-soliton compression could be initiated when both the driving pulse energy and gas pressure inside the hollow waveguide were properly adjusted [4]. This high-order-soliton compression process leads to simultaneously strong spectral broadening of the driving pulse, and as the short-wavelength tail of the broadened spectrum reaches certain phase-matched wavelength, high-efficiency energy transfer from the driving to the phase-matched wavelength can be obtained, resulting in a packet of quasi-linear wave (named as DW) shedding from the driving pulse [4,5]. The phase-matching condition of this emission process strongly relies on the dispersion landscape of the gas-filled hollow waveguide, which can be manipulated through adjusting the gas type and pressure in the hollow waveguide, rendering highly-flexible wavelength tunability of the generated DW, from visible to vacuum ultraviolet [1-3,6-8].

Different from anti-resonant hollow-core fiber (ARHCF) with tens of micrometers core size [6-8], HCF has generally a core diameter of several hundreds of micrometers so as to support low-loss pulse propagation [1-3]. The large core size naturally leads to low nonlinearity which increases the pulse energy capacity of the capillary waveguide. In order to obtain high-order solitons, the pulse energy launched into the HCF should normally be tens to hundreds of μJ [1-3], which is one to two orders of magnitudes higher than that needed for small-core ARHCFs [6-8], scaling up the pulse energy of the short-wavelength DW to several-μJ level [1-3]. Another remarkable advantage of this technique is that it can be used, in principle, to generate high-energy, few-fs ultraviolet pulses, being very important in applications that need high temporal resolution, such as ultrafast spectroscopy and molecular dynamics probing [9,10]. Even though few-fs temporal widths of DW pulses have long been predicted in theory, direct access of µJ-level few-fs ultraviolet pulses generated from gas-filled hollow-core fibers, has proved difficult to achieve. Ermolov et al. measured 275 nm DW pulses with tens of nJ pulse energies from a Ne-filled kagomé-style ARHCF [11], using a transient-grating cross-correlation frequency-resolved optical gating (TG-XFROG) set-up. The long pulse duration of ~280 fs at the measurement location was mainly caused by the large positive chirp induced by the pulse propagation in air and optical elements [11]. Using an in-vacuum XFROG set-up, Brahms et al. measured directly ~3 fs DW pulses at wavelengths from 225 nm to 300 nm generated in an Ar-filled kagomé-style ARHCF [12], however the relatively-low pulse energy (~30 nJ) as well as rather complicated set-up with vacuum chamber and pressure gradient limit to some extent its practical applications.

In this letter, we demonstrated comprehensive studies both in theory and in the experiment on the temporal profile of µJ-level ultraviolet DW pulses generated in a short length of Ar-filled HCF. The results exhibit that the accumulated pulse chirp outside the HCF can be properly compensated using several chirped mirrors, and µJ-level few-fs pulses at 384 nm and 430 nm were obtained and measured in atmosphere environment using a home-built self-diffraction frequency-resolved optical gating (SD-FROG) set-up. In addition, the results show that the ultraviolet pulse width is rather sensitive to both the driving pulse energy and the HCF length, providing a few useful clues for future system design.

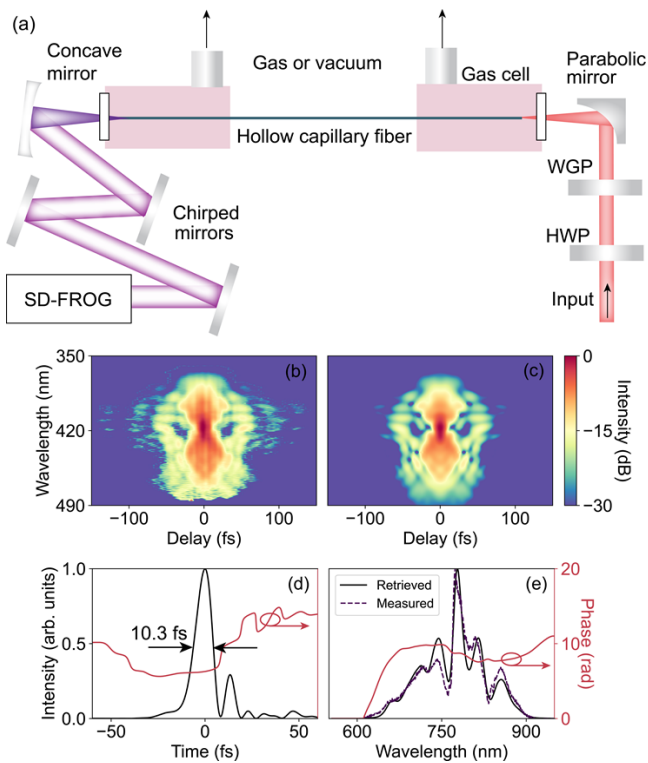

Fig. 1. (a) Experimental set-up. HWP, half-wave plate; WGP, wire grid polarizer. (b-e) SHG-FROG traces of the driving pulses at 800 nm, giving a FWHM pulse width of 10.3 fs with small residual chirp.

The set-up used in the experiment is illustrated in Fig. 1(a). The 800 nm pulses from a commercial Ti:Sapphire ultrafast laser system at a repetition rate of 1 kHz, pass through a pulse-compression stage which is composed of a section of 78-cm-long, 250-µm-core-diameter Ar-filled silica capillary together with several chirped mirrors to compensate the pulse chirp [13,14]. The pulse-compression stage can deliver 800 nm pulses with a maximum pulse energy of ~173 µJ and a pulse duration of ~10 fs. As shown in Fig. 1(a), the energy of the driving pulse can be adjusted using a half-wave plate and a wire grid polarizer, and a second-harmonic-generation frequency-resolved optical gating (SHG-FROG) was used to measure its temporal profile, giving a full-width-half-maximum (FWHM) pulse width of 10.3 fs with negligible residual pulse chirp [see Figs. 1(b)-1(e)]. The driving pulses were then coupled into the Ar-filled HCF using a parabolic mirror with a focal length of 30 cm. The HCF has a length of 35 cm and a core diameter of 100 µm. As we gradually decreased the Ar-gas pressure inside the HCF from ~1.33 bar to ~0.28 bar, the phase-matched wavelength of DW emission was shifted from near-ultraviolet to deep-ultraviolet. In the experiment, we increased the driving pulse energy and observed µJ-energy DW pulses out of the HCF with the central wavelength tuned from ~430 nm to ~210 nm. Three typical output spectra from the HCF are illustrated in Fig. 2, exhibiting emission spectra with DW wavelengths at 350 nm, 380 nm and 430 nm, when the gas pressures and launched driving pulse energies were (38 µJ, 976 mbar), (30 µJ, 1064 mbar) and (30 µJ, 1331 mbar) respectively.

The generated ultraviolet pulses were characterized using the SD-FROG technique. In the experimental set-up, several ultraviolet chirped mirrors were used to compensate the positive chirp induced by the in-air pulse propagation and the $MgF_2$ window mounted on the gas cell [see Fig. 1(a)].

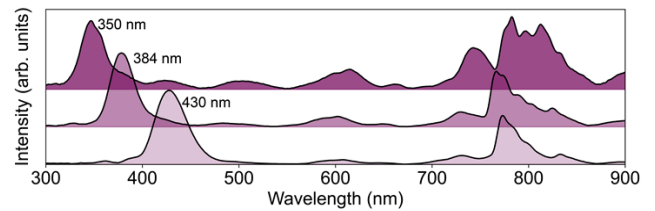

Fig. 2. Measured spectra of the output light from the HCF. The DW wavelengths are at 350 nm, 384 nm and 430 nm, corresponding to 976 mbar, 1064 mbar and 1331 mbar Ar-gas pressure in the HCF.

In the experiment, we selected three chirped mirrors of the same type with high reflectivity over the spectral range of 350 nm to 450 nm, which were also used to filter out the residual driving pulse. The group delay dispersion (GDD) provided by each mirror is -37 $fs^2$, and the free-space length in the set-up was measured to be ~1.85 m, giving rise to total in-air GDD of 87.8 $fs^2$ [15]. The 0.5-mm-thick $MgF_2$ window of the gas cell also provides positive GDD of 26.4 $fs^2$ [15], then the additional GDD generated outside the HCF is compensated to be 3.2 $fs^2$ as summarized in Table 1. This careful compensation, on one hand, ensures that the SD-FROG results can well reflect the temporal profiles of the DW at the output port of the HCF, on the other hand it ensures ultraviolet pulses with few-fs widths obtained at the measurement location.

**Table. 1 Dispersion values of different elements at 384 nm**

| Dispersion source | GDD (fs$^2$) |
|---|---|
| 0.5-mm-thick MgF$_2$ | 26.4[a] |
| 3 chirped mirrors | -111.0 |
| 1.85-m-long air | 87.8[a] |
| Total | 3.2 |

[a]Reference [15].

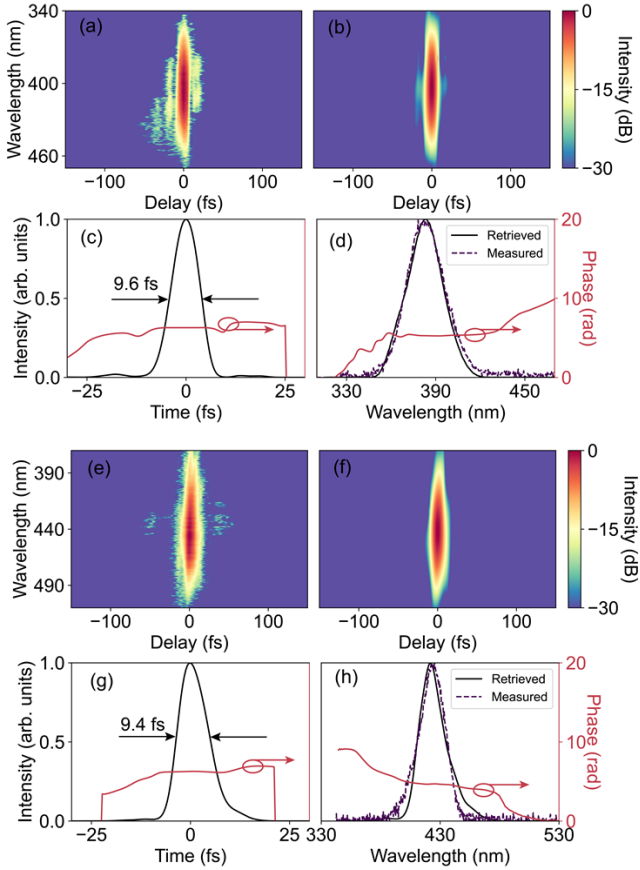

Fig. 3. SD-FROG measurements of DW pulses at 384 nm (a-d) and 430 nm (e-h). (a, e) Measured FROG traces. (b, f) Retrieved FROG traces. (c, d, g, h) Retrieved temporal and spectral profiles and phases of the pulses. The pulse spectra measured directly using spectrometer are plotted as purple-dashed lines in (d, h).

The SD-FROG results of the 384 nm and 430 nm DW pulses are illustrated in Fig. 3. At both of the two wavelengths, the measured SD-FROG traces [see Figs. 3(a) and 3(e)] show good agreement with the retrieved traces [see Figs. 3(b) and 3(f)] with a small FROG error of ~0.3%, and the retrieved spectra [as black-solid lines in Figs. 3(d) and 3(h)] agree well with the directly-measured spectra [as purple-dashed lines in Figs. 3(d) and 3(h)] using the optical spectrometer. The retrieved pulse profiles are plotted in Figs. 3(c) and 3(g), exhibiting short pulse widths of 9.6 fs at 384 nm and 9.4 fs at 430 nm, close to the Fourier-transform limits (8.2 fs at 384 nm and 7.5 fs at 430 nm). It is shown in Figs. 3(a) and 3(e) that the shapes of FROG traces for both two cases are essentially vertical, indicating the weak residual chirp. We also retrieved the phase information of the measured pulses, giving residual GDD values of 3.5 fs$^2$ and 4.4 fs$^2$ at 384 nm and 430 nm. In order to test the accuracy of the chirp measurement, we removed the chirped mirrors one by one in the experiment, and found that the retrieved pulse GDD increased almost linearly by 37 fs$^2$ per mirror.

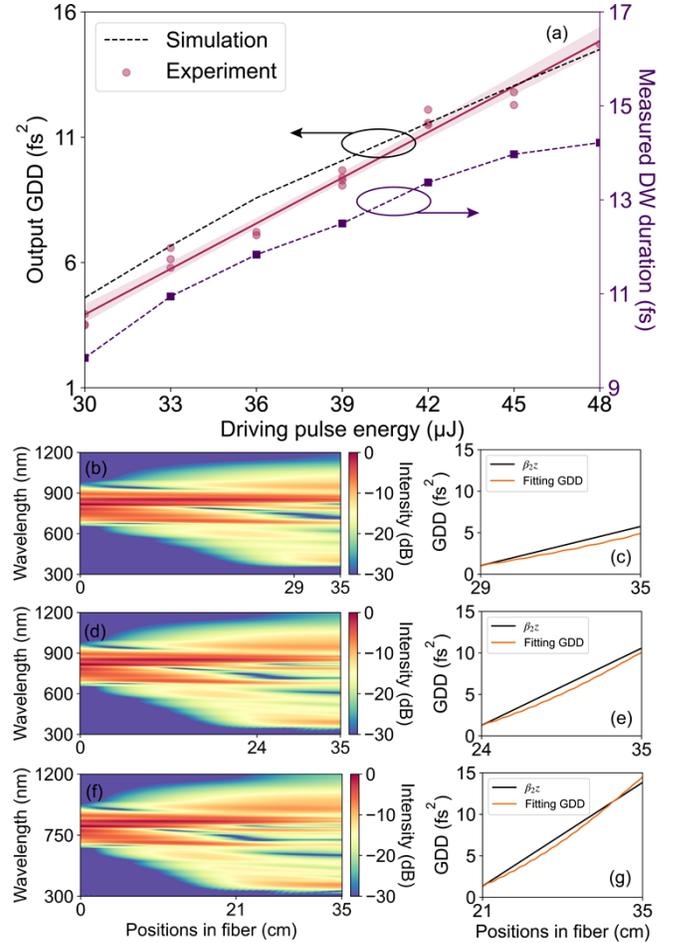

Fig. 4. (a) Simulated and measured pulse GDD and width at 384 nm, plotted as the function of the driving pulse energy. (b, d, f) Simulated spectral evolutions in the 35-cm-long HCF at different driving pulse energies. (c, e, g) Simulated GDDs (as orange lines) of DW pulses accumulated in the HCF, obtained using the simulation results. Pulse GDDs calculated simply using the second-order dispersion of the HCF are also plotted as black lines.

Using this set-up, we further studied the influence of the driving pulse energy and HCF length on the pulse chirp and width of the ultraviolet DW. When the Ar-gas pressure in the HCF was fixed at 1064 mbar, we gradually increased the driving pulse energy from 30 µJ to 48 µJ, corresponding to an increase of the soliton order from 3.9 to 4.9. We measured the SD-FROG traces at different driving pulse energies and retrieved both pulse chirp and FWHM width of the DW pulses. The data are summarized in Fig. 4(a). It can be observed that as the driving pulse energy increased, the GDD of the 384 nm DW pulse increased almost linearly from 3.5 fs$^2$ to 14.7 fs$^2$, leading to an increase of the pulse width from 9.6 fs to 14.2 fs. This phenomenon can be explained as: when gradually increasing the driving pulse energy in the HCF, the resulting stronger nonlinearity accelerated the high-order-soliton compression process, leading to the moving-forward of in-HCF position where the ultraviolet DW starts to emit. This can result in a longer propagation distance of the DW in the HCF and therefore larger positive chirp of the output ultraviolet pulses.

The experimental observations were perfectly reproduced in our theoretical modeling. Using the unidirectional pulse propagation model [16], we simulated the nonlinear pulse propagation and short-wavelength DW emission process, the simulation results are illustrated in Figs. 4(b)-4(g). It is shown that as the driving pulse energy increased from 30 μJ to 39 μJ and then to 48 μJ, the position of the DW generation in the HCF moves forward from ~29 cm to ~24 cm and then to ~21 cm [see Figs. 4(b), 4(d) and 4(f)]. The simulation results also show that the DW at its generation position in the HCF has a small GDD value of ~1 fs$^2$ which is insensitive to the driving pulse energy. After generation, the positive chirp of the DW accumulates linearly over propagation in the HCF, mainly due to second-order dispersion of the capillary waveguide, highlighting its essential characteristic of quasi-linear wave, see Figs. 4(c), 4(e) and 4(g). These results indicate that in order to obtain high-energy, transform-limit DW pulses out of the HCF, the system parameters need to be carefully adjusted so as to make the HCF length a little longer than the length needed for high-efficiency DW emission, and the optimized HCF length is strongly dependent on the nonlinear dynamics in the HCF, being sensitive to driving pulse energy, gas pressure and DW wavelength.

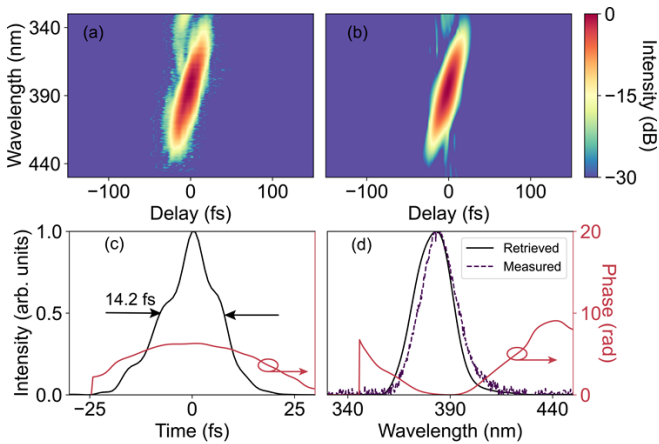

Fig. 5. SD-FROG results of the 384 nm DW pulses at the driving pulse energy of 48 μJ. The retrieved pulse width is 14.2 fs, and some temporal pulse distortion can be clearly observed.

Even though the additional GDD value induced by the increase of driving pulse energy (from 30 μJ to 48 μJ) is not large (~11 fs$^2$), its influence on the pulse width and shape of the ultraviolet DW pulse is not negligible. As shown in Fig. 5, the measured SD-FROG trace of 384 nm DW pulse at driving pulse energy of 48 μJ gives a FWHM pulse with of 14.2 fs, with a distorted pulse temporal profile [see Fig. 5(c)].

In conclusion, we studied temporal information of ultraviolet DW pulses generated in a 35-cm-long Ar-filled HCF. Using three ultraviolet chirped mirrors, precise chirp compensation was implemented, leading to the direct measurement of μJ-level, sub-10-fs ultraviolet DW pulses at 384 nm and 430 nm in atmosphere environment. Both theoretical and experimental studies reveal that energy variation of the driving pulse can vary the nonlinear dynamics of DW emission, leading to additional linear chirp and therefore broadening and distortion of the pulse temporal profile. The measurement range of this scheme is currently limited by the reflection window of the chirped mirrors which can be extended to the deep or even vacuum ultraviolet regime. Our results reveal some useful insight into the dispersive-wave-emission dynamics in HCFs and the set-up demonstrated can provide wavelength-tunable, μJ-level ultraviolet pulses with few-fs pulse widths.

**Funding.** Zhangjiang Laboratory Construction and Operation Project (20DZ2210300); National Postdoctoral Program for Innovative Talents (BX2021328); China Postdoctoral Science Foundation (2021M703325); Shanghai Science and Technology Innovation Action Plan (21ZR148270); National High-level Talent Youth Project; National Key R&D Program of China (2017YFE0123700); National Natural Science Foundation of China (61925507); Strategic Priority Research Program of the Chinese Academy of Sciences (XDB1603).

†These authors contributed equally to this letter.

### References

1. J. C. Travers, T. F. Grigorova, C. Brahms, and F. Belli, "High-energy pulse self-compression and ultraviolet generation through soliton dynamics in hollow capillary fibres," Nat. Photonics **13**(8), 547–554 (2019).
2. C. Brahms, T. Grigorova, F. Belli, and J. C. Travers, "High-energy ultraviolet dispersive-wave emission in compact hollow capillary systems," Opt. Lett. **44**(12), 2990–2993 (2019).
3. C. Brahms, F. Belli, and J. C. Travers, "Infrared attosecond field transients and UV to IR few-femtosecond pulses generated by high-energy soliton self-compression," Phys. Rev. Res. **2**(4), 043037 (2020).
4. J. C. Travers, W. Chang, J. Nold, N. Y. Joly, and P. St.J. Russell, "Ultrafast nonlinear optics in gas-filled hollow-core photonic crystal fibers [Invited]," J. Opt. Soc. Am. B **28**(12), A11–A26 (2011).
5. C. Markos, J. C. Travers, A. Abdolvand, B. J. Eggleton, and O. Bang, "Hybrid photonic-crystal fiber," Rev. Mod. Phys. **89**(4), 045003 (2017).
6. N. Y. Joly, J. Nold, W. Chang, P. Holzer, A. Nazarkin, G. K. Wong, F. Biancalana, and P. S. Russell, "Bright spatially coherent wavelength-tunable deep-UV laser source using an Ar-filled photonic crystal fiber," Phys. Rev. Lett. **106**(20), 203901 (2011).
7. A. Ermolov, K. F. Mak, M. H. Frosz, J. C. Travers, and P. S. J. Russell, "Supercontinuum generation in the vacuum ultraviolet through dispersive-wave and soliton-plasma interaction in a noble-gas-filled hollow-core photonic crystal fiber," Phys. Rev. A **92**(3), 033821 (2015).
8. F. Belli, A. Abdolvand, W. Chang, J. C. Travers, and Philip St. J. Russell, "Vacuum-ultraviolet to infrared supercontinuum in hydrogen-filled photonic crystal fiber," Optica **2**(4), 292-300 (2015).
9. A. H. Zewail, "Femtochemistry: atomic-scale dynamics of the chemical bond using ultrafast lasers (Nobel Lecture)," Angew. Chem. Int. Ed. Engl. **39**(15), 2586-2631 (2000).
10. A. Stolow, A. E. Bragg, and D. M. Neumark, "Femtosecond time-resolved photoelectron spectroscopy," Chem. Rev. **104**(4), 1719-1757 (2004).
11. A. Ermolov, H. Valtna-Lukner, J. C. Travers, and P. St.J. Russell, "Characterization of few-fs deep-UV dispersive waves by ultra-broadband transient-grating XFROG," Opt. Lett. **41**(23), 5535–5538 (2016).
12. C. Brahms, D. R. Austin, F. Tani, A. S. Johnson, D. Garratt, J. C. Travers, J. W. G. Tisch, P. St.J. Russell, and J. P. Marangos, "Direct characterization of tuneable few-femtosecond dispersive-wave pulses in the deep UV," Opt. Lett. **44**(4), 731–734 (2019).
13. Z. Y. Huang, Y. F. Chen, F. Yu, D. Wang, R. R. Zhao, Y. Zhao, S. F. Gao, Y. Y. Wang, P. Wang, M. Pang, and Y. X. Leng, "Continuously wavelength-tunable blueshifting soliton generated in gas-filled photonic crystal fibers," Opt. Lett. **44**(7), 1805-1808 (2019).
14. Y. Chen, Z. Huang, F. Yu, D. Wu, J. Fu, D. Wang, M. Pang, Y. Leng, and Z. Xu, "Photoionization-assisted, high-efficiency emission of a dispersive wave in gas-filled hollow-core photonic crystal fibers," Opt. Express **28**(11), 17076-17085 (2020).
15. M. N. Polyanskiy, "Refractive index database," https://refractiveindex.info/ Accessed on 2022-05-21.
16. Z. Y. Huang, D. Wang, Y. F. Chen, R. R. Zhao, Y. Zhao, S. Nam, C. Lim, Y. J. Peng, J. Du, and Y. X. Leng, "Wavelength-tunable few-cycle pulses in visible region generated through soliton-plasma interactions," Opt. Express **26**(26), 34977–34993 (2018).